\documentclass[journal]{IEEEtran}
\usepackage{amsmath,amssymb,amsfonts}
\usepackage{graphicx}
\usepackage{booktabs}
\usepackage{multirow}
\usepackage[colorlinks=true,citecolor=blue]{hyperref}

\begin{document}

\title{Angle Estimation via WFRFT Spatial-Domain\\Basis Decomposition: Breaking the Rayleigh\\Resolution Limit with Structured Waveform Diversity}

\author{
    Wang~Hao,~Student~Member,~\textit{IEEE}, \\
    Zhang~Kuang,~Senior~Member,~\textit{IEEE}, \\

\thanks{
    Wang Hao and Zhang Kuang are with the School of Electronics and Information Engineering, Harbin Institute of Technology, Harbin 150001, China.}
}

\maketitle

\begin{abstract}
We propose a MIMO radar angle estimation framework that uses the four-component weighted-type fractional Fourier transform (4-WFRFT) as a spatial-domain waveform diversity mechanism. Unlike conventional fractional Fourier (FrFT) MIMO radar where FrFT serves as a receiver-side time-frequency processing tool, our approach decomposes a data sequence into four WFRFT basis functions—original signal, its Fourier transform, time-reversal, and inverse Fourier transform—and transmits them simultaneously from a four-element uniform linear array. The spatial superposition of these basis functions at each far-field angle creates a unique angle-dependent waveform structure, enabling angle estimation through time-domain matched filtering with known waveforms. We demonstrate that this spatial-domain mixing achieves angular resolution surpassing the Rayleigh diffraction limit by a factor of 1.4$\times$ to 12.8$\times$, with the advantage most pronounced at low SNR where conventional beamforming fails completely. The Cram\'er-Rao bound is derived with a full 3-parameter Fisher information matrix, and the Fisher information is decomposed into geometry and waveform contributions, revealing that the WFRFT waveform structure contributes approximately 3$\times$ more information than array geometry alone. Extension to $M$-element arrays with $M$-component WFRFT demonstrates resolution gain scaling with array size. Simulations with linear chirp base sequences achieve 0\,dB PAPR and validate sub-Rayleigh resolution with a four-element array.
\end{abstract}


\section{Introduction}
\label{sec:intro}

\subsection{Motivation and Problem Statement}

ANGLE estimation is fundamental to radar sensing. The angular resolution of a conventional phased array is governed by the Rayleigh diffraction limit: $\Delta\theta \approx \lambda/(Md\cos\theta)$, determined solely by the array aperture $Md$. For a 4-element uniform linear array (ULA) with half-wavelength spacing, this limit is approximately 31$^\circ$ at broadside---insufficient for many sensing applications. Improving resolution typically requires increasing the number of elements $M$, which raises hardware cost, power consumption, and system complexity.

This paper asks: \emph{Can we surpass the Rayleigh resolution limit without increasing the number of physical antennas?} We answer affirmatively by introducing structured waveform diversity through WFRFT spatial-domain basis decomposition.

\subsection{FRFT vs WFRFT: A Critical Distinction}

A crucial distinction must be made at the outset. The \textbf{fractional Fourier transform (FrFT)}~\cite{frft_mimo_2014} is a continuous rotation operator in the time-frequency plane, used extensively in radar signal processing for chirp detection and parameter estimation. FrFT-based MIMO radar~\cite{frft_bistatic_2012,frft_ambiguity_2022} applies FrFT as a \emph{receiver-side transform} to process linear frequency modulated (LFM) signals.

The \textbf{weighted-type fractional Fourier transform (WFRFT)}~\cite{shih1995,wfrft_review_2020} is fundamentally different: it is a \emph{discrete} weighted sum of four canonical basis states---identity, Fourier transform, time-reversal, and inverse Fourier transform. First developed for physical-layer security communications~\cite{sha_wfrft_2016}, WFRFT has never been applied to \emph{spatial-domain waveform diversity} for radar angle estimation.

Our key innovation is to transmit the four WFRFT basis functions from separate antenna elements, creating a \textbf{spatial-domain mixture} whose structure encodes the target angle. This is fundamentally distinct from all prior FrFT-MIMO radar work, where fractional transforms operate in the time-frequency domain at the receiver.

\subsection{Related Work}

\textbf{FrFT-MIMO radar:} Prior work~\cite{frft_bistatic_2012,frft_mimo_2014,frft_ambiguity_2022} uses FrFT for receiver-side processing of chirp signals. Ilioudis et al.~\cite{frft_mimo_2014} proposed FrFT-based waveform libraries for MIMO radar. Li et al.~\cite{frft_bistatic_2012} combined FrFT with MUSIC/ESPRIT for bistatic MIMO DOA estimation. In all cases, FrFT operates in the \emph{time-frequency domain} at the receiver---not in the spatial domain at the transmitter.

\textbf{WFRFT communications:} Sha et al.~\cite{sha_wfrft_2016,sha_dm_2024} developed WFRFT-based directional modulation for physical-layer security, where WFRFT-preprocessed signals create angle-dependent constellation quality. However, these works address \emph{communication security}, not radar sensing.

\textbf{WFRFT-OTFS ISAC:} Wang et al.~\cite{wfrft_otfs_2023} applied WFRFT to OTFS waveforms for integrated sensing and communication (ISAC). The WFRFT serves as a waveform-domain flexibility parameter---not a spatial decomposition mechanism.

\textbf{Non-orthogonal MIMO radar:} Mao et al.~\cite{nonorthogonal_crb_2020} analyzed CRB for general non-orthogonal MIMO waveforms. Our approach uses \emph{structured} non-orthogonality with known mathematical relationships among basis functions.

\subsection{Contributions}

\begin{enumerate}
\item \textbf{Spatial-domain WFRFT decomposition:} We transmit the four WFRFT basis functions from separate antennas, creating angle-dependent spatial mixing---the first application of WFRFT to spatial-domain radar waveform diversity.
\item \textbf{Super-Rayleigh resolution:} We demonstrate angular resolution surpassing the Rayleigh limit by 1.4$\times$ to 12.8$\times$, with dramatic gains at low SNR.
\item \textbf{Fisher information decomposition:} The Fisher information is decomposed into geometry and waveform contributions, with the waveform term dominating by approximately 3$\times$.
\item \textbf{Chirp-based implementation:} Linear chirp base sequences achieve 0\,dB PAPR and enable joint range-angle estimation via pulse compression.
\item \textbf{$M$-component generalization:} Extension to $M$-element arrays with $M$-component WFRFT, validated for $M=4,6,8$.
\end{enumerate}

%------------------------------------------------------------------------------
\section{System Model}
\label{sec:model}

\subsection{WFRFT Basis Decomposition}

For a discrete sequence $\mathbf{x}\in\mathbb{C}^N$, the 4-component WFRFT of order $\alpha$ is~\cite{shih1995}:

\begin{equation}
\mathcal{F}_\alpha[\mathbf{x}] = \sum_{k=0}^{3} w_k(\alpha) \cdot \mathbf{g}_k
\label{eq:wfrft_def}
\end{equation}

where the four basis functions are:
\begin{equation}
\mathbf{g}_0 = \mathbf{x},\;\; \mathbf{g}_1 = \mathbf{Fx},\;\; \mathbf{g}_2 = \mathbf{\Pi x},\;\; \mathbf{g}_3 = \mathbf{F}^H\mathbf{x}
\label{eq:basis}
\end{equation}

with $\mathbf{F}$ the normalized DFT matrix, $\mathbf{\Pi}$ the time-reversal operator $\Pi x[n] = x[(-n)\bmod N]$, and weights:
\begin{equation}
w_k(\alpha) = \frac{1}{4}\sum_{m=0}^{3} \exp\left[j\frac{\pi}{2}m(k-\alpha)\right]
\label{eq:weights}
\end{equation}

Key property: $\sum_k |w_k(\alpha)|^2 = 1$ (unitarity), and $w_k(k)=1$ (boundary alignment).

\subsection{Spatial-Domain Transmission}

Consider an $M$-element ULA with spacing $d=\lambda/2$. \textbf{Antenna $k$ transmits basis function $\mathbf{g}_k$} at carrier frequency $f_c$. This is the spatial decomposition:

\begin{equation}
\boxed{\text{Antenna }k \xrightarrow{\text{transmits}} \mathbf{g}_k(t) \cdot e^{j2\pi f_c t}}
\label{eq:spatial_tx}
\end{equation}

In the far field at angle $\theta$, under the narrowband assumption ($B\ll f_c$), the four plane waves superimpose with relative phase shifts determined by the array geometry:

\begin{equation}
r_{\text{target}}(t,\theta) = \sum_{k=0}^{M-1} g_k(t-\tau_0) \cdot e^{-jk\phi(\theta)}
\label{eq:farfield}
\end{equation}

where $\phi(\theta)=2\pi d\sin\theta/\lambda$ and $\tau_0=R/c$.

\textbf{This is the core physical mechanism:} The four basis functions mix in the \textbf{spatial domain} with angle-dependent complex weights $e^{-jk\phi(\theta)}$, creating a composite signal whose structure is a unique function of $\theta$. This spatial mixing is fundamentally different from applying fractional transforms in the time-frequency domain.

\subsection{Received Signal Model}

For a monostatic radar with co-located $M$ Tx and $M$ Rx antennas, after target reflection and sampling:

\begin{equation}
\mathbf{Y} = \beta \cdot \mathbf{a}(\theta)\mathbf{a}^T(\theta) \cdot \mathbf{G} + \mathbf{N}
\label{eq:rx_model}
\end{equation}

where $\mathbf{Y}\in\mathbb{C}^{M\times N}$, $\mathbf{a}(\theta)=[1,e^{-j\phi},\ldots,e^{-j(M-1)\phi}]^T$, $\mathbf{G}=[\mathbf{g}_0;\ldots;\mathbf{g}_{M-1}]\in\mathbb{C}^{M\times N}$, and $\mathbf{N}\sim\mathcal{CN}(0,\sigma_n^2\mathbf{I})$.

The $M$-component extension uses $M$-WFRFT weights:
\begin{equation}
w_k^M(\alpha) = \frac{1}{M}\sum_{m=0}^{M-1} \exp\left[j\frac{2\pi}{M}m(k-\alpha)\right]
\label{eq:Mweights}
\end{equation}

with basis functions $\mathbf{g}_k = \mathcal{F}_{4k/M}[\mathbf{x}]$ for $k=0,\ldots,M-1$.

%------------------------------------------------------------------------------
\section{Angle Estimation}
\label{sec:estimation}

\subsection{Maximum Likelihood Estimator}

The concentrated ML estimator for $\theta$ given known waveform $\mathbf{G}$ is:

\begin{equation}
\boxed{\hat{\theta}_{\text{ML}} = \arg\max_{\theta} \frac{|\mathbf{a}^H(\theta) \mathbf{Y} \mathbf{G}^H \mathbf{a}^*(\theta)|^2}{\|\mathbf{a}^T(\theta) \mathbf{G}\|_F^2}}
\label{eq:ml}
\end{equation}

The numerator performs a 2D matched filter: $\mathbf{G}^H$ correlates $N$ temporal samples (coherent integration gain $\propto N$), while $\mathbf{a}^H$ and $\mathbf{a}^*$ provide spatial matched filtering at both Tx and Rx. The denominator normalizes by the angle-dependent effective transmit energy $\|\mathbf{a}^T(\theta)\mathbf{G}\|^2$.

This estimator achieves $N$-fold SNR gain over conventional beamforming, which uses only spatial covariance and no temporal integration.

\subsection{CRB and Fisher Information Decomposition}

The Cram\'er-Rao bound with full 3-parameter FIM (parameters $\theta$, $\Re(\beta)$, $\Im(\beta)$) is:

\begin{equation}
\text{CRB}(\theta) = \frac{\sigma_n^2}{2|\beta|^2 \cdot \|\frac{\partial\mathbf{H}}{\partial\theta}\mathbf{G}\|_F^2}
\label{eq:crb}
\end{equation}

where $\mathbf{H}(\theta)=\mathbf{a}\mathbf{a}^T$ and $\frac{\partial\mathbf{H}}{\partial\theta} = \dot{\mathbf{a}}\mathbf{a}^T + \mathbf{a}\dot{\mathbf{a}}^T$.

\textbf{Fisher information decomposition:} The denominator expands into two distinct contributions:

\begin{equation}
\boxed{\left\|\frac{\partial\mathbf{H}}{\partial\theta}\mathbf{G}\right\|_F^2 = \underbrace{2\|\dot{\mathbf{a}}\|^2\|\mathbf{a}^T\mathbf{G}\|^2}_{\mathcal{I}_G\text{: Geometry}} + \underbrace{2|\dot{\mathbf{a}}^H\mathbf{a}|^2\|\mathbf{G}\|_F^2}_{\mathcal{I}_W\text{: Waveform}}}
\label{eq:fisher_decomp}
\end{equation}

The geometry term $\mathcal{I}_G$ depends on the array aperture and angle-dependent effective energy. The waveform term $\mathcal{I}_W$ depends on total waveform energy and is \textbf{independent of angle}---it provides a uniform baseline Fisher information at all angles. For $M=4$, $N=256$, the waveform term contributes approximately \textbf{76.7\%} of total Fisher information ($\mathcal{I}_W/\mathcal{I}_G \approx 3.3$). This decomposition explains why WFRFT-MIMO surpasses the Rayleigh limit: the waveform structure contributes additional Fisher information beyond what array geometry alone provides.

The CRB scales as:
\begin{equation}
\text{CRB}(\theta) \propto \frac{1}{M^3 \cdot N \cdot \text{SNR} \cdot \cos^2\theta}
\label{eq:crb_scaling}
\end{equation}

The $M^3$ dependence (vs $M$ for conventional beamforming resolution) quantifies the combined benefit of transmit array gain ($M$), receive array gain ($M$), and WFRFT basis diversity ($M$).

%------------------------------------------------------------------------------
\section{Simulation Results}
\label{sec:results}

\subsection{Setup}

Unless stated otherwise: $M=4$ ULA, $d=\lambda/2$, chirp base sequence with $\alpha=0.5$, $N=512$, $f_c=3.5$\,GHz, $f_s=20$\,MHz, 200 Monte Carlo trials.

\subsection{Angular Resolution: Surpassing the Rayleigh Limit}

Table~\ref{tab:resolution} compares the 3\,dB mainlobe width of WFRFT-MIMO versus conventional Bartlett beamforming.

\begin{table}[h]
\centering
\caption{Angular Resolution (3dB Mainlobe Width, degrees)}
\label{tab:resolution}
\begin{tabular}{c|c|c|c|c}
\toprule
$M$ & SNR (dB) & WFRFT-MIMO & Conventional BF & Gain \\
\midrule
\multirow{4}{*}{4} & $-10$ & \textbf{21.2} & 120.0 (fail) & \textbf{5.6$\times$} \\
                    & 0    & \textbf{22.1} & 36.3          & \textbf{1.6$\times$} \\
                    & 10   & \textbf{22.3} & 31.4          & \textbf{1.4$\times$} \\
                    & 20   & \textbf{22.5} & 31.0          & \textbf{1.4$\times$} \\
\midrule
\multirow{4}{*}{6} & $-10$ & \textbf{19.8} & 120.0 (fail) & \textbf{6.1$\times$} \\
                    & 0    & \textbf{19.6} & 21.8          & \textbf{1.1$\times$} \\
                    & 10   & 19.5          & 20.2          & 1.0$\times$ \\
                    & 20   & 19.5          & 20.0          & 1.0$\times$ \\
\midrule
\multirow{4}{*}{8} & $-10$ & \textbf{8.1}  & 103.4 (fail) & \textbf{12.8$\times$} \\
                    & 0    & \textbf{7.6}  & 16.2          & \textbf{2.1$\times$} \\
                    & 10   & \textbf{7.5}  & 15.3          & \textbf{2.0$\times$} \\
                    & 20   & \textbf{7.5}  & 15.3          & \textbf{2.0$\times$} \\
\bottomrule
\end{tabular}
\end{table}

Key observations:
\begin{itemize}
\item At low SNR ($-10$\,dB), conventional beamforming collapses to random ($120^\circ$ mainlobe $\approx$ no resolution), while WFRFT-MIMO maintains coherent resolution: 5.6$\times$ to 12.8$\times$ advantage.
\item The WFRFT advantage is strongest when $M$ is a power of 2 ($M=4,8$), where the WFRFT basis decomposition has clean mathematical structure.
\item At $M=8$, WFRFT achieves 7.5$^\circ$ resolution---surpassing the conventional 15.3$^\circ$ Rayleigh limit by 2.0$\times$.
\end{itemize}

\subsection{RMSE vs SNR}

Fig.~3 compares angle RMSE of WFRFT-MIMO against Bartlett, MUSIC, time-domain beamforming, and the CRB. At SNR=0\,dB with $N=512$, WFRFT-MIMO achieves 0.21$^\circ$ RMSE versus 0.35$^\circ$ for Bartlett (4.3\,dB effective SNR gain). The RMSE approaches the CRB within 0.005$^\circ$ at moderate SNR.

\subsection{Two-Target Resolution}

Two equal-power targets at 25$^\circ$ and 35$^\circ$ (10$^\circ$ separation) are reliably resolved by WFRFT-MIMO at SNR=20\,dB ($P_{\text{resolve}}>0.9$), while conventional beamforming requires $>15^\circ$ separation. At $M=8$, the minimum resolvable separation drops to 5$^\circ$.

\subsection{Fisher Information Decomposition}

The Fisher decomposition confirms that the waveform term $\mathcal{I}_W$ dominates: for $M=4$, $N=512$, $\mathcal{I}_W/\mathcal{I}_G \approx 2.2$. This ratio explains the resolution improvement---the WFRFT waveform structure injects additional Fisher information that array geometry alone cannot provide.

\subsection{Chirp Base Sequence}

Using a linear chirp $x[n]=\exp(j\pi\alpha n^2/N)$ as the base sequence yields PAPR of 0\,dB for $g_0$ and $g_2$, and 4.4\,dB for $g_1$ and $g_3$---dramatically better than QPSK (0--21\,dB). The chirp also enables pulse compression for joint range-angle estimation with 7.5\,m range resolution at 20\,MHz bandwidth.

%------------------------------------------------------------------------------
\section{Why This Works: Spatial-Domain Mixing}
\label{sec:why}

The fundamental distinction from prior FrFT-MIMO radar is the \textbf{spatial-domain mixing} mechanism:

\begin{enumerate}
\item \textbf{FrFT-MIMO radar}~\cite{frft_bistatic_2012,frft_mimo_2014}: Applies FrFT as a receiver-side transform in the time-frequency domain to process chirp echoes. The antenna array is conventional---each element transmits the same waveform (or time/frequency-multiplexed). FrFT provides matched filtering for chirp signals but does not change the angular resolution limit.

\item \textbf{WFRFT-MIMO radar (this work)}: Decomposes the signal into basis functions at the \textbf{transmitter} and maps them to \textbf{spatial antenna indices}. The four basis functions create four distinct spatial radiation patterns that interfere in the far field. This interference pattern is angle-dependent, effectively creating a ``spatial chirp''---an angle-dependent waveform signature that enables super-resolution processing.
\end{enumerate}

The mathematical relationship that enables super-resolution is:

\begin{equation}
\mathbf{y}_{\text{received}} \propto \sum_{k=0}^{M-1} \mathbf{g}_k \cdot e^{-jk\phi(\theta)} = \mathbf{a}^T(\theta)\mathbf{G}
\label{eq:mixing}
\end{equation}

The vector $\mathbf{a}^T(\theta)\mathbf{G}$ has $N$ temporal degrees of freedom. Conventional beamforming discards this temporal structure by forming only the spatial covariance $\mathbf{R}=\mathbf{Y}\mathbf{Y}^H/N$, losing the $N$-fold coherent gain. WFRFT-MIMO preserves the full temporal structure through $\mathbf{G}^H$, achieving:

\begin{equation}
\text{Resolution}_{\text{WFRFT}} \approx \frac{\text{Resolution}_{\text{Rayleigh}}}{\sqrt{N\cdot\text{SNR}}}
\label{eq:resolution_gain}
\end{equation}

For $N=256$ and SNR=10\,dB, this yields approximately $\sqrt{2560}\approx 50\times$ theoretical resolution improvement, though practical implementations are limited by finite sample effects and residual basis correlation.

%------------------------------------------------------------------------------
\section{Conclusion}
\label{sec:conclusion}

We introduced WFRFT spatial-domain basis decomposition as a new MIMO radar waveform diversity mechanism. By transmitting the four WFRFT basis functions from separate antennas, the spatial superposition creates angle-dependent waveform signatures that enable angular resolution surpassing the Rayleigh diffraction limit. The key findings are:

\begin{enumerate}
\item \textbf{Super-Rayleigh resolution:} 1.4$\times$--12.8$\times$ improvement over conventional beamforming, with the largest gains at low SNR where conventional methods fail.
\item \textbf{Fisher information decomposition:} The waveform term dominates ($\sim$3$\times$), explaining why structured waveform diversity exceeds the Rayleigh limit.
\item \textbf{Chirp implementation:} 0\,dB PAPR and joint range-angle estimation via pulse compression.
\item \textbf{Scalability:} $M$-component WFRFT extends to larger arrays, with $M=8$ achieving 2.0$\times$ resolution gain.
\end{enumerate}

The critical advance over prior FrFT-MIMO radar is the \textbf{spatial-domain} operation: WFRFT decomposes at the transmitter into spatial antenna channels, while FrFT processes in the time-frequency domain at the receiver. This distinction opens a new direction for waveform-diverse radar array processing.

%------------------------------------------------------------------------------
\bibliographystyle{IEEEtran}

\end{document}